\documentclass[12pt,a4paper]{article}
  
\usepackage{graphicx}
\usepackage{amsmath}  
\usepackage{amsfonts,amssymb}
\usepackage{subfigure}
\usepackage{bm}

\usepackage[utf8x]{inputenc}
\usepackage[margin=25mm]{geometry}
\usepackage[numbers]{natbib}
\usepackage{theorem}
\usepackage{float}

\def\Iset{{\mathbb I}}
\def\Eset{\textnormal{E}}
\def\weib{\displaystyle \exp\left\{\hspace*{-0.08cm} -\left(\hspace*{-0.08cm}\frac{t_i}{\eta_j} \hspace*{-0.08cm}\right)^{\beta_j} \hspace*{-0.08cm}\right\}}

{\theorembodyfont{\rmfamily} \newtheorem{example}{Example}}

\begin{document}
\thispagestyle{empty}

\begin{center}
{\Huge Reliability estimators for the components of series and parallel systems: The Weibull model}

\vspace{0.3cm}
{\bf Felipe L. Bhering$^1$, Carlos Alberto de Bragan\c{c}a Pereira$^1$, Adriano Polpo$^2$\let\thefootnote\relax\footnote{Address for correspondence: Adriano Polpo, Universidade Federal de S\~{a}o Carlos, Departamento de Estat\'{i}stica, Rod. Washington Luiz, km 235, CEP: 13565-905, S\~{a}o Carlos/SP, Brazil. E-mail: polpo@ufscar.br}
}\\
$^1$ Department of Statistics, University of S\~{a}o Paulo, Brazil\\
$^2$ Department of Statistics, Federal University of S\~{a}o Carlos, Brazil\\

{\large Feb 13, 2013}
\end{center}

\begin{abstract} 
This paper presents a hierarchical Bayesian approach to the estimation of components' reliability (survival) using a Weibull model for each of them. The proposed method can be used to estimation with general survival censored data, because the estimation of a component's reliability in a series (parallel) system is equivalent to the estimation of its survival function with right- (left-) censored data. Besides the Weibull parametric model for reliability data, independent gamma distributions are considered at the first hierarchical level for the Weibull parameters and independent uniform distributions over the real line as priors for the parameters of the gammas. In order to evaluate the model, an example and a simulation study are discussed.  
\end{abstract}


\section{Introduction}
This paper presents a hierarchical Bayesian approach to the estimation of components' reliability using a Weibull model for each component in series and parallel systems. A series system is a frame of components that works if and only if all its components are functional, that is, whenever one fails the system fails. As a dual frame, the parallel system fails if and only if all components are malfunctioning.

The literature dealing with the problem of estimating the reliability of series systems, or competing risks, is abundant. The work of \citet{KaplanMeier1958} is arguably the most celebrated work, where it was developed a nonparametric estimator using a frequentist approach. For a Bayesian counterpart, we draw the attention to \citet{Salinas2002} and \citet{Polpo2011}. \citet{Rodrigues2012} performed a simulation study of three different methods to estimate the reliability of a series system. They compared the Kaplan-Meier estimator \cite{KaplanMeier1958}, maximum likelihood estimator (MLE) and the Bayesian plug-in estimator (BPE) for Weibull reliability systems. Their results indicated that MLE and BPE are similar in quality and that both outperformed the Kaplan-Meier estimator. However, the construction of credible bounds was not addressed in their work.

For parallel systems, the literature is scarce. To the best of our knowledge, \citet{Polpo2009} were the first to address the nonparametric reliability estimation in parallel systems and their components, using the Bayesian paradigm. Later on, \citet{Bhattacharya2010} proposed a frequentist nonparametric estimator for components' reliability under a restrictive condition that all components are independent and identically distributed.

In a related work, \citet{Polpo2009b} presented the reliability estimation with Weibull models and non-informative priors, also using the Bayesian paradigm. Their proposal had very demanding computational needs, and the described Markov Chain Monte Carlo (MCMC) algorithm suffered from convergence issues in many problem instances, making the estimation a difficult task. The authors realized later on that it is often impossible to evaluate the components' reliability with that method because of such issues, and unfortunately one cannot predict when the algorithm will succeed. The present work aims at achieving a robust estimation procedure for the estimation problem. We suggest to consider non-informative priors in an one-level hierarchical model as means to solve the estimation issues faced by the algorithm of \citet{Polpo2009b}. Moreover, an important goal of this work is to provide a simple way to build credible bounds for the reliability functions, giving a step-by-step algorithm that performs such analysis. 

Similarly to \citet{Polpo2009b} and \citet{Rodrigues2012}, we consider Weibull statistical models for the system's reliability. Two independent gamma distributions are considered in the first hierarchical level. The gamma distributions are parametrized by their means and variances, instead of the more common parametrization with scale and shape. In the second level of the hierarchy, we choose two flat priors for the means, and fixed values for the variances of the gamma distributions in the first level. These hyper-parameters corresponding to variances in the first level can be seen as prior precision parameters. The means of the gamma distributions (first level of the hierarchy) can be viewed as the prior expectations of the parameters of the Weibull model. In this work, the posterior modes are taken as the Bayesian estimators of the gamma distributions.

The estimation of the reliability functions has three main steps: (i) we draw a sample from the posterior distribution of the Weibull parameters; (ii) using the appropriate transformation, we build a sample from the reliability posterior distribution; and (iii) locally, for each reliability time, we evaluate the posterior mean. The high posterior density (HPD) procedure was used to define the credible region for the reliability function. We emphasize that we are not using the plug-in estimator, but the posterior mean of the reliability function, which seems more suitable under the Bayesian paradigm.

This paper is organized as follows. Section \ref{sec:model} describes all functions that are involved in the estimation procedure. Section \ref{sec:est} provides the estimation procedure itself. Section \ref{sec:ex} presents a simulation study that highlights the quality of the model and the proposed estimators, and final remarks and additional comments are given in Section \ref{sec:final}. We note that an extended abstract of this work has appeared in the Brazilian Conference on Bayesian Statistics \citep{Bhering2012a}.

\section{The model}
\label{sec:model}

We use the same notation as in \citet{Polpo2009b}. Consider a system of $k$ components and let $X_j$, $j \in \{1,\ldots, k\}$, be the sequence of failure times of all components. We assume that this sequence is composed of independent random variables with a (possibly) different Weibull distribution for each component. Recall that we only observe a random vector of two variables, namely, $(T,\delta)$ with $T=\min(X_1, \ldots, X_k)$ for the series system and $T=\max(X_1, \ldots, X_k)$ for the parallel system, with $\delta=j$ if $T = X_j$, for $j=1, \ldots, k$. The $\delta$ quantity can be viewed as an indicator function of the component that caused the system failure. 

Consider a sample of $n$ independent and identically distributed systems (either all series or all parallel systems). The observations are represented by $(T,\delta)=\{(T_i, \delta_i): i=1, \ldots, n\}$. The reliability function of the $j$th component is given by $R_j(t) = P(X_j > t)$, $j=1,\ldots, k$. Therefore, the reliability function is $R(t) = \prod_{j=1}^{k} R_j(t)$ for the series system and $R(t)= 1-\prod_{j=1}^{k}(1-R_j(t))$ for the parallel system.

We define random variables $X_j$ for the components' reliability with Weibull distributions parametrized by $\theta_j = (\beta_j, \eta_j)$, that is,
\begin{equation}
P(X_j > x | \theta_j) = R(x | \theta_j) = \exp\left\{-\left({\frac{x}{\eta_j}}\right)^{\beta_j} \right\}
\end{equation}
for $x > 0$, $\beta_j > 0$ (shape) and $\eta_j > 0$ (scale). Then, the likelihood function for the series system is given by
\begin{equation}
L(\bm{\theta}|t,\delta)\propto \prod_{j=1}^{k}\prod_{i=1}^{n} [f_j(t_i|\theta_j)]^{\Iset_{\{\delta_i=j\}}} [R_j(t_i|\theta_j)]^{1-\Iset_{\{\delta_i=j\}}},
\end{equation}
and for the parallel system,
\begin{equation}
L(\bm{\theta}|t,\delta)\propto \prod_{j=1}^{k}\prod_{i=1}^{n} [f_j(t_i|\theta_j)]^{\Iset_{\{\delta_i=j\}}} [1-R_j(t_i|\theta_j)]^{1-\Iset_{\{\delta_i=j\}}},
\end{equation}
where $f$ is the density function of a random variable with Weibull distribution, $\bm{\theta} = (\theta_1, \ldots, \theta_k)$, and $\Iset_A$ is the indicator function of the set $A$.

The prior distributions were considered independent with $\beta_j \sim$ gamma($m_{\beta_j}, v_{\beta_j}$), $\eta_j \sim$ gamma($m_{\eta_j}, v_{\eta_j}$), $\pi(m_{\beta_j}) \propto \pi(m_{\eta_j}) \propto 1$, and $v_{\beta_j}$ and $v_{\eta_j}$ as known constants, $j = 1, \ldots, k$. Then
\begin{eqnarray*}
\pi(\bm{\vartheta}) & \propto & \pi(\bm{\theta} | \bm{m_\beta}, \bm{v_\beta}, \bm{m_\eta}, \bm{v_\eta}) \pi(\bm{m_\beta}) \pi(\bm{v_\beta}) \pi(\bm{m_\eta}) \pi(\bm{v_\eta}) \\
& \propto & \prod_{j=1}^{k} \pi(\theta_j | m_{\beta_j}, v_{\beta_j}, m_{\eta_j}, v_{\eta_j}) \pi(m_{\beta_j}) \pi(v_{\beta_j}) \pi(m_{\eta_j}) \pi(v_{\eta_j}) \\
& \propto & \prod_{j=1}^{k} \frac{\beta_j^{m_{\beta_j}^2/v_{\beta_j}-1} \exp\{ -m_{\beta_j} \beta_j/v_{\beta_j} \} }{(v_{\beta_j}/m_{\beta_j})^{m_{\beta_j}^2/v_{\beta_j}} \Gamma(m_{\beta_j}^2/v_{\beta_j})} \frac{\eta_j^{m_{\eta_j}^2/v_{\eta_j}-1} \exp\{ -m_{\eta_j} \eta_j/v_{\eta_j} \} }{(v_{\eta_j}/m_{\eta_j})^{m_{\eta_j}^2/v_{\eta_j}} \Gamma(m_{\eta_j}^2/v_{\eta_j})},
\end{eqnarray*}
where $\bm{\vartheta} = (\bm{\theta}, \bm{m_\beta}, \bm{v_\beta}, \bm{m_\eta}, \bm{v_\eta})$,  $\bm{m_\beta} = (m_{\beta_1}, \ldots, m_{\beta_k})$ and $\bm{m_\eta} = (m_{\eta_1}, \ldots, m_{\eta_k})$ are the prior mean parameters, $\bm{v_\beta} = (v_{\beta_1}, \ldots, v_{\beta_k})$ and $\bm{v_\eta} = (v_{\eta_1}, \ldots, v_{\eta_k})$ are the prior variance (precision) parameters, and $m_{\beta_j}, m_{\eta_j}, v_{\beta_j}, v_{\eta_j} > 0$, $j= 1, \ldots, k$.

In this case, we have that the posterior distributions of series and parallel systems are, respectively,
\begin{equation*}
\pi(\bm{\vartheta}|\bm{t},\bm{\delta})\propto \pi(\bm{\vartheta}) \prod_{i=1}^{n} {\left[\frac{{t_i}^{\beta_j-1} \beta_j }{\eta_j}  \weib \right]}^{\Iset_{\{ \delta_i=j \}}} {\left[\weib \right]}^{1-\Iset_{\{ \delta_i=j \}}},
\end{equation*}
and
\begin{equation*}
\pi(\bm{\vartheta}|\bm{t},\bm{\delta})\propto \pi(\bm{\vartheta}) \prod_{i=1}^{n} {\left[\frac{{t_i}^{\beta_j-1} \beta_j }{\eta_j}  \weib \right]}^{\Iset_{\{ \delta_i=j \}}} {\left[1-\weib \right]}^{1-\Iset_{\{ \delta_i=j \}}},
\end{equation*}
where $\bm{t} = (t_1, \ldots, t_n)$ are the observed failure time of the system, $\bm{\delta} = (\delta_1, \ldots, \delta_n)$ are the indicators of which component failed, and the other quantities are as defined before.

\section{Estimation}
\label{sec:est}

For the estimation, we use the EM algorithm \citep{McLachlan2008} to obtain the posterior mode as estimates of $\bm{m_\beta}$ and $\bm{m_\eta}$, and the MCMC procedure to generate a sample from the posterior distribution of $\bm{\beta}$ and $\bm{\eta}$. The algorithm steps for the estimation are briefly given as follows:

\begin{enumerate}
\item Choose the prior precision values $\bm{v_\beta}$ and $\bm{v_\eta}$. Note that one can set to the same value all the precision values, that is, $v = \bm{v_\beta} = \bm{v_\eta}$. We suggest the use of $v = 4$.

\item Choose the initial guess for the parameters to be estimated: $\bm{m_\beta}$, $\bm{m_\eta}$, $\bm{\beta}$ and $\bm{\eta}$.

\item \label{MCMC} Using the initial guess, consider $\bm{m_\beta}$ and $\bm{m_\eta}$ as fixed values. Employing the MCMC tool, generate a sample (of size $n_p$) from the posterior distribution of $\bm{\beta}$ and $\bm{\eta}$. We suggest the use of $n_p = 1000$. It may also be necessary to use a ``burn-in'' and a ``jump'' to ensure convergence of the MCMC.

\item \label{EM1} (Expectation step of the EM) Using the posterior sample of $\bm{\beta}$ and $\bm{\eta}$ obtained in Step \ref{MCMC}, evaluate the mean of the likelihood function, obtaining a ``mean'' function of $\bm{m_\beta}$ and $\bm{m_\eta}$.

\item \label{EM2} (Maximization step of the EM) Find the values of $\bm{m_\beta}$ and $\bm{m_\eta}$ that maximize the function obtained in Step \ref{EM1}.

\item \label{last} Update the initial guess of $\bm{m_\beta}$ and  $\bm{m_\eta}$ with the values obtained in Step \ref{EM2}, and the values of $\bm{\beta}$ and $\bm{\eta}$ with their posterior mean obtained in Step \ref{MCMC}.

\item Repeat Steps \ref{MCMC}, \ref{EM1}, \ref{EM2} and \ref{last} until convergence of $\bm{m_\beta}$ and $\bm{m_\eta}$ is reached. We suggest the use of a tolerance value of three decimal places between the previous values of $\bm{m_\beta}$ and $\bm{m_\eta}$ and the current ones, in order to decide whether to stop iterating.

\item Once convergence of $\bm{m_\beta}$ and $\bm{m_\eta}$ is reached, use their values to generate a sample from the posterior distribution of $\bm{\beta}$ and $\bm{\eta}$ by applying the MCMC tool.
\end{enumerate}

Using this algorithm, we obtain the posterior mode (the Bayesian estimate) of $\bm{m_\beta}$ and $\bm{m_\eta}$, and a sample (of size $n_p$) from the joint posterior distribution of $\bm{\beta}$ and $\bm{\eta}$. 
If we estimate the reliability function of any component (let us arbitrarily choose component 1), then we can notice that the estimations of the other components' reliability functions are very similar and could be omitted here. Consider that the sample from the posterior of the parameters of component 1 can be expressed as $(\beta_{11}, \beta_{12}, \ldots, \beta_{1n_p})$ and $(\eta_{11}, \eta_{12}, \ldots, \eta_{1n_p})$. To obtain the reliability estimates and credible regions, consider the functions $Y_\ell(t) = F(t \mid \beta_{1\ell}, \eta_{1\ell})$, $\ell = 1, \ldots, n_p$, where $F(t \mid \beta_{1\ell}, \eta_{1\ell})$ are Weibull distribution functions conditioned on $\beta_{1\ell}$ and $\eta_{1\ell}$. Consequently, the posterior mean estimate of the component's reliability can be expressed as
\[ \widehat{R(t)} = \Eset[R(t) \mid data] = 1- \frac{1}{n_p} \sum\limits_{\ell=1}^{n_p} Y_\ell(t), ~\textnormal{for each} ~t. \]

Hence, for each fixed $t$, $Y(t) = (Y_1(t), \ldots, Y_{n_p}(t))$ is a sample from the posterior of the component 1's distribution function and, to obtain the credible region, we can either use the quantiles of $Y(t)$ or evaluate the high posterior density credible interval. To estimate the mean reliability time, we have
\[ \Eset[T \mid data] = \frac{1}{n_p} \sum\limits_{\ell=1}^{n_p} \Eset[T \mid \beta_{1\ell}, \eta_{1\ell}], ~\ell = 1, \ldots, n_p, \]
where $\Eset[T \mid \beta_{1\ell}, \eta_{1\ell}] = \eta_{1\ell} \Gamma(1 + 1/\beta_{1\ell})$ is the mean of a random variable with Weibull distribution, and $\Gamma(\cdot)$ is the gamma function. Note that similar procedures can be used to evaluate other quantities of interest; as an example, the posterior median could be evaluated.

\section{Examples}
\label{sec:ex}

\begin{example} ~\\
\label{ex_1}
Consider three random variables $X_1,X_2,X_3$ such that $X_1$ has Weibull distribution with mean $2$ and variance $4$, $X_2$ has gamma distribution with mean $2$ and variance $0.667$, and $X_3$ has log-normal distribution with mean $2.014$ and variance $6.968$. We have generated a sample (with size $n=100$) of series systems with these three components and another sample (again with size $n=100$) of parallel systems with the exactly same three components. The components were chosen in order to have similar means but different variances and, consequently, different distributions. We have used the same theoretical components in both simulations (series and parallel systems) to verify, in each situation, the differences that are due to the distinct system models with the available data. The simulated data have the following characteristics: (i) for the series systems, we have obtained 64\%, 80\%, and 56\% of censured data for components 1, 2, and 3, respectively; and (ii) for the parallel systems, we have observed 61\%, 68\% and 71\%, respectively for the same three components. In this case, the main interest is in the estimation of the components' reliability functions. Note that, with our simulated example, we have a huge amount of censored data, making it a challenging example.

As already said, the estimation procedures are performed using MCMC. We have discarded the first $10,000$ samples (as {\it burn-in}) from the posterior to achieve the stationary measure and then have generated a sample from the posterior. To perform the estimation of the reliability functions and the credible region, we have used a sample of size $1,000$ from the posterior, which was obtained by discarding $10$ samples (the {\it jump} between each final sample point). We have used $v = 4$ in the prior specification for all parameters and systems. For the experiment with series systems, we have obtained $\widehat{\bm{m_\beta}} = (1.31, 4.12, 1.44)$ and $\widehat{\bm{m_\eta}} = (2.19, 2.03, 1.80)$. For the parallel systems, the estimates are $\widehat{\bm{m_\beta}} = (1.28, 2.57, 0.92)$ and $\widehat{\bm{m_\eta}} = (2.67, 2.28, 2.11)$. To evaluate the quality of these estimates, we have compared the ``true'' reliability of each component with the estimated reliability function. Table \ref{tab_1} presents the posterior mean and the posterior standard deviation of each parameter involved in the model (for both series and parallel systems). Note that the standard deviations are relatively small, and the estimation of the mean reliability time is very close to the original values, indicating a good performance of our method.

\begin{table}[H]
\centering
\caption{\label{tab_1} Posterior mean (sd) of some quantities involved in the estimation of the simulated examples.}
\begin{tabular}{cccc}
\hline
            & \multicolumn{3}{c}{Series system estimates}\\
            & $\beta$     & $\eta$      & $\Eset(T \mid \beta,\eta)$ \\
\hline
Component 1 & 1.26 (0.17) & 2.27 (0.41) & 2.13 (0.45) \\
Component 2 & 3.98 (0.53) & 2.06 (0.12) & 1.87 (0.11) \\
Component 3 & 1.40 (0.17) & 1.83 (0.22) & 1.68 (0.22) \\
\hline
            & \multicolumn{3}{c}{Parallel system estimates} \\
            & $\beta$     & $\eta$      & $\Eset(T \mid \beta,\eta)$ \\
\hline
Component 1 & 1.23 (0.15) & 2.60 (0.27) & 2.45 (0.22) \\
Component 2 & 2.47 (0.32) & 2.25 (0.14) & 2.00 (0.13) \\
Component 3 & 0.87 (0.11) & 1.98 (0.33) & 2.17 (0.29) \\
\hline
\end{tabular}
\end{table}

It can be seen from the 95\% credible bounds that the ``true'' reliability of each component was well estimated. We note however that the ``true'' reliability functions of the components are, for short (time) intervals, outside the 95\% credible bounds. Considering that these are reasonably challenging examples, this situation is likely to happen in any estimation procedure (see Figures \ref{fig:series} and \ref{fig:parallel}).

\begin{figure}[H]
\centering
\subfigure{\includegraphics[scale=0.35,keepaspectratio=true]{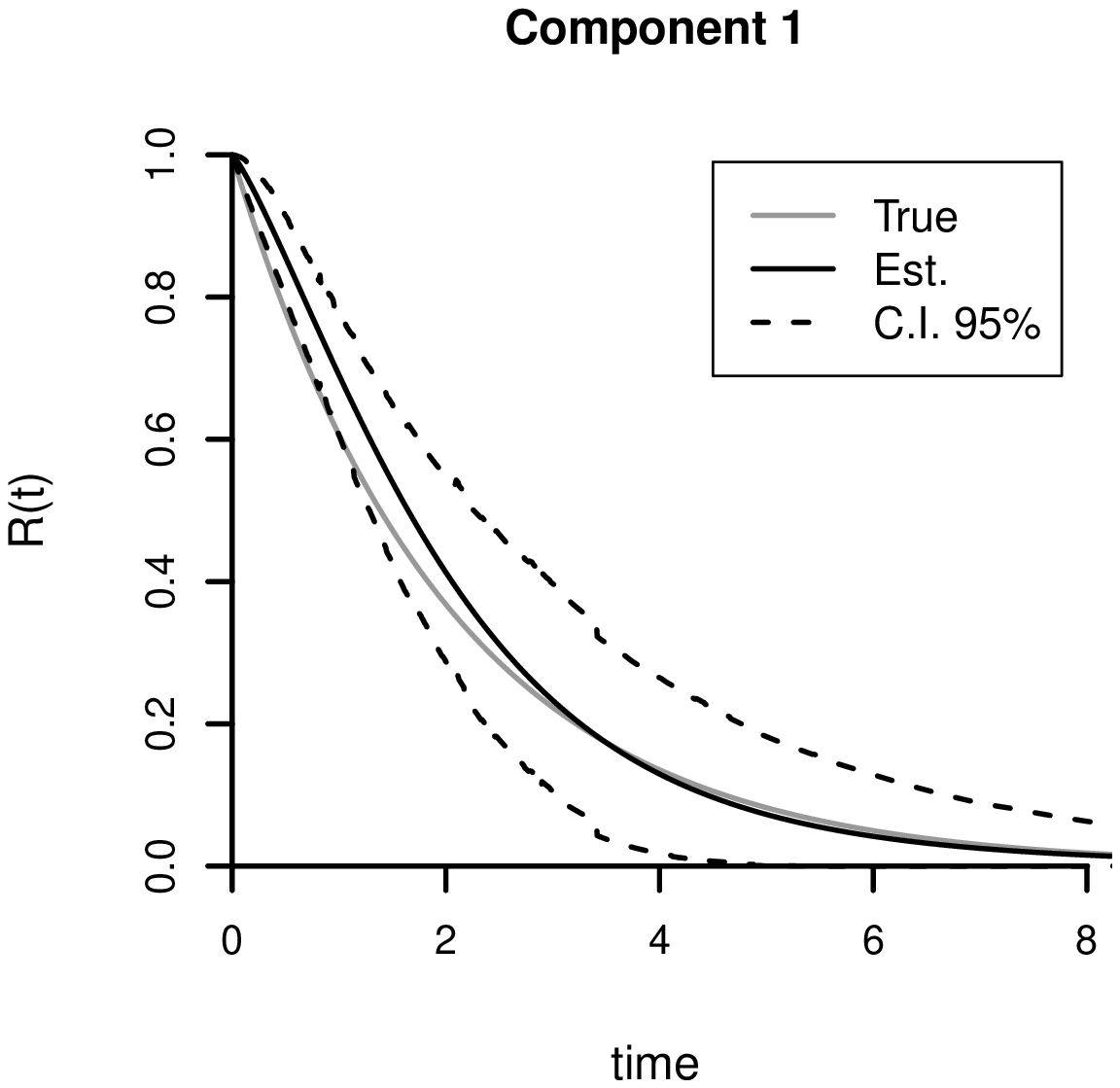}
\label{fig_1a}}
\qquad
\subfigure{\includegraphics[scale=0.35,keepaspectratio=true]{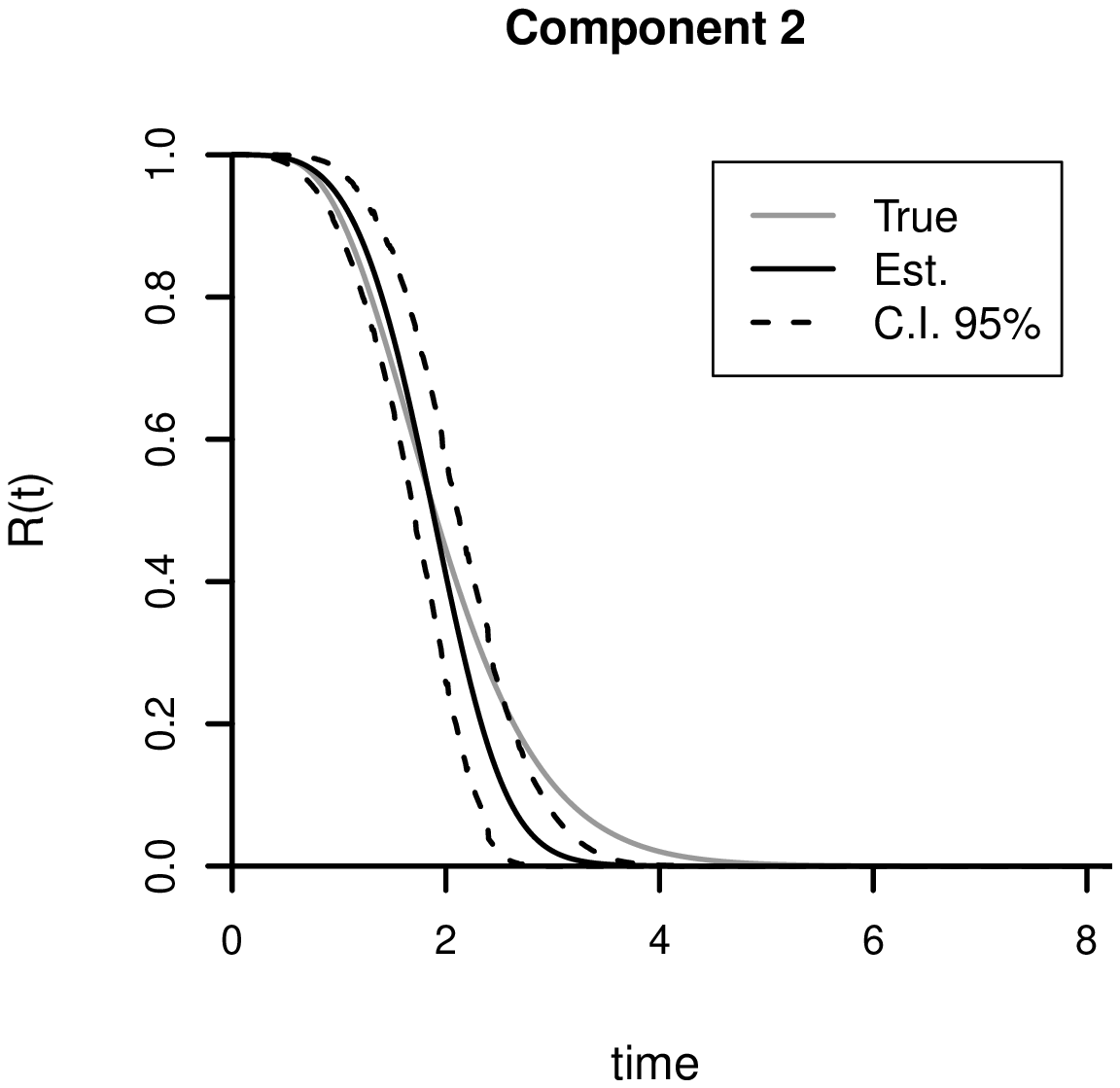}
\label{fig_1b}}
\qquad
\subfigure{\includegraphics[scale=0.35,keepaspectratio=true]{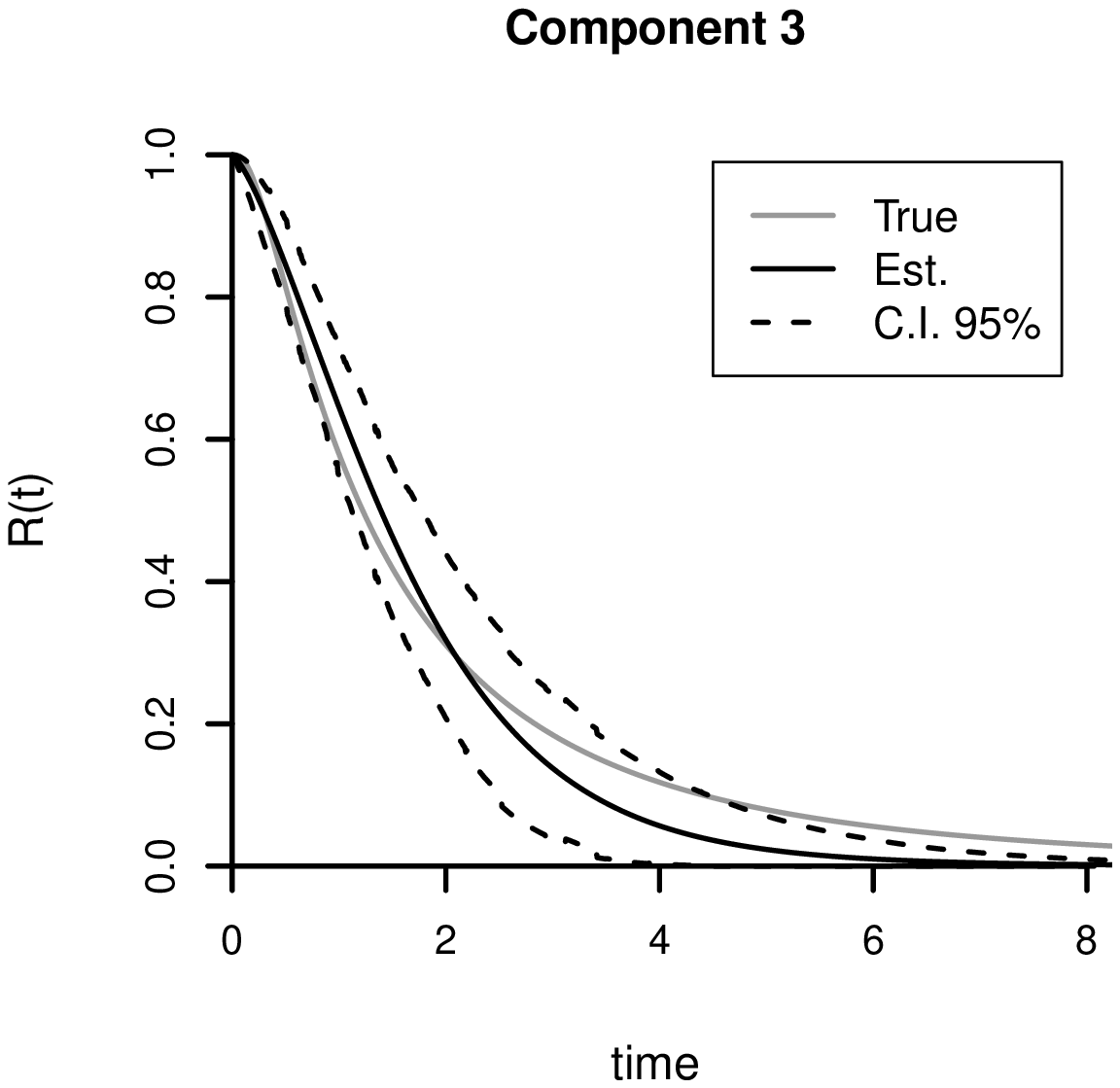}
\label{fig_1c}}
\caption{\label{fig:series} Reliability of the components in the experiment with series systems:
         \subref{fig_1a} Component 1; \subref{fig_1b} Component 2; \subref{fig_1c} Component 3.}
\end{figure}

\begin{figure}[H]
\centering
\subfigure{\includegraphics[scale=0.35,keepaspectratio=true]{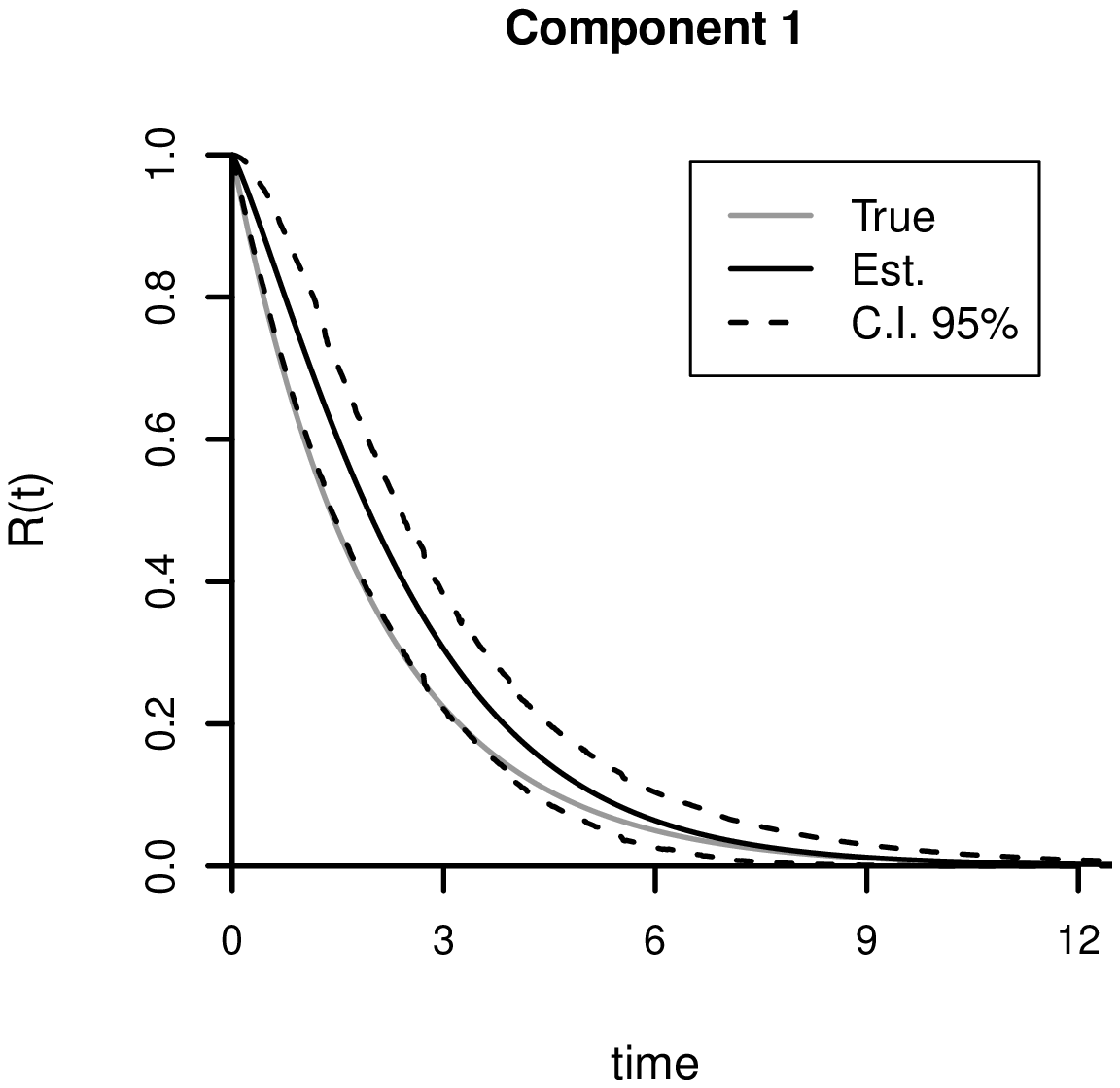}
\label{fig_2a}}
\qquad
\subfigure{\includegraphics[scale=0.35,keepaspectratio=true]{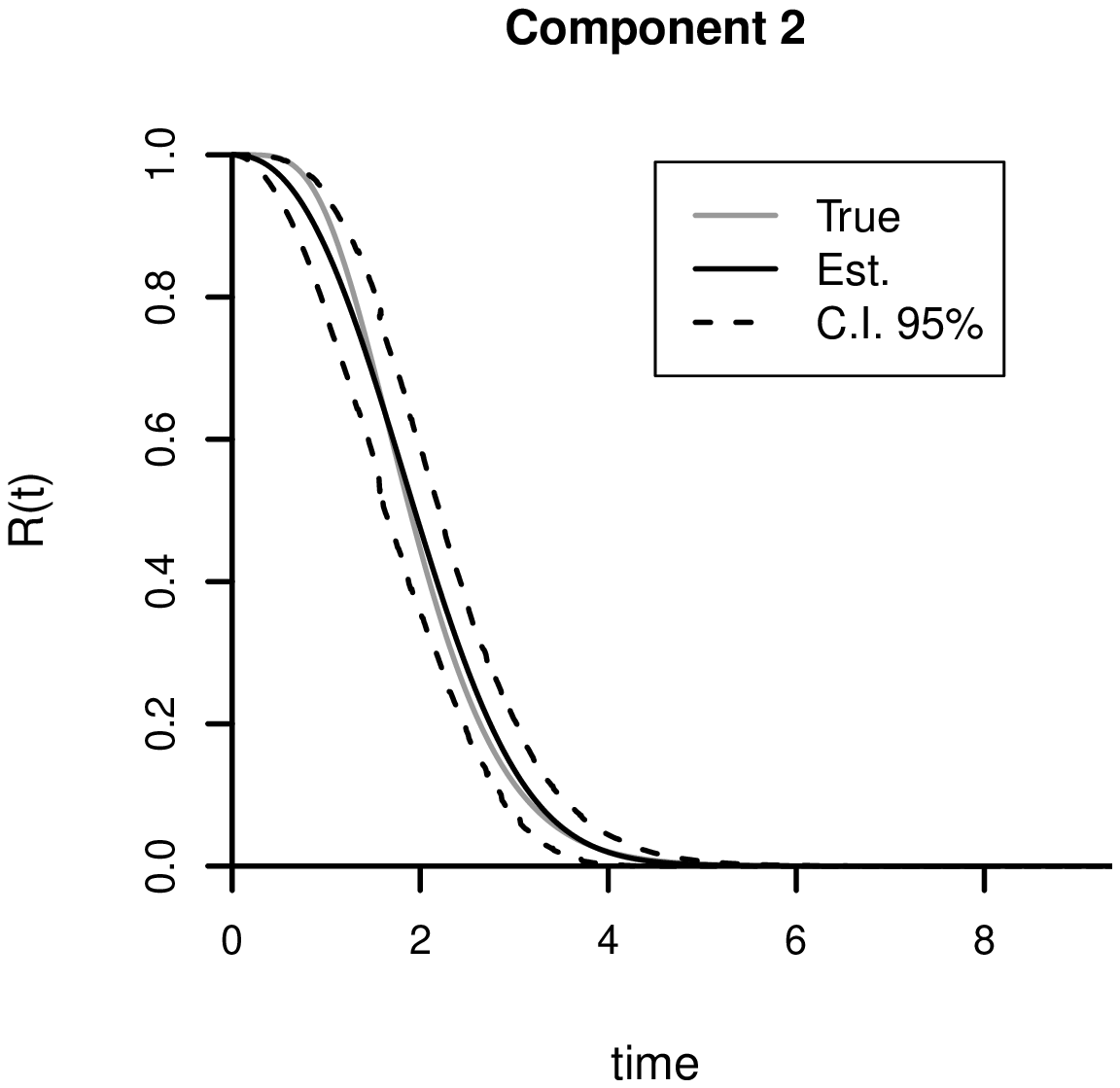}
\label{fig_2b}}
\qquad
\subfigure{\includegraphics[scale=0.35,keepaspectratio=true]{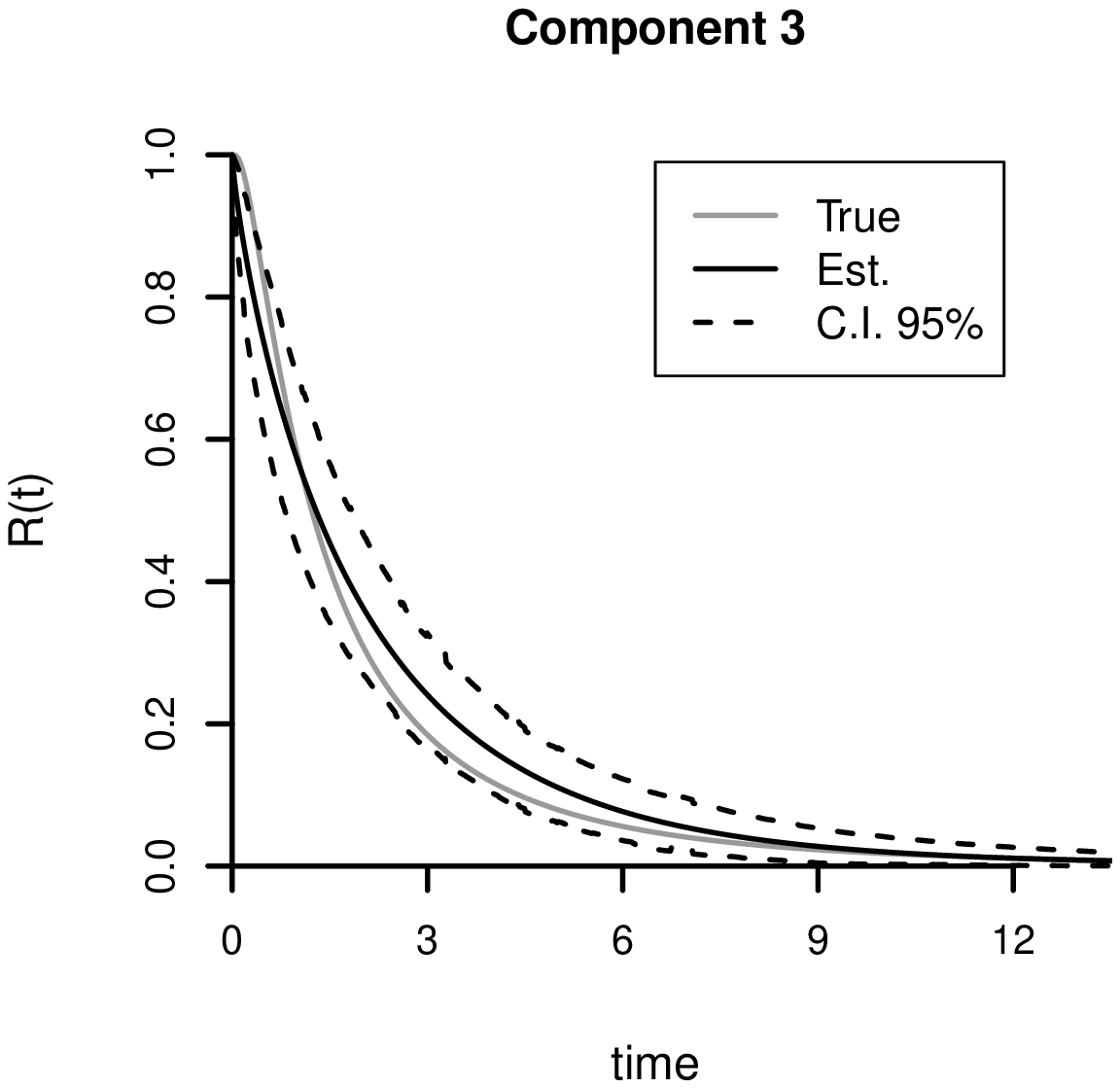}
\label{fig_2c}}
\caption{\label{fig:parallel} Reliability of the components in the experiment with parallel systems:
         \subref{fig_1a} Component 1; \subref{fig_1b} Component 2; \subref{fig_1c} Component 3.}
\end{figure}
\end{example}


\begin{example} ~\\
\label{ex_2}
This example has a simulation study to show the quality of the proposed hierarchical Weibull model over many different conditions.
We have considered $108$ different scenarios that were built using three different sample sizes ($n = 30, 100, 1000$), three different proportions of censored data ($0\%$, $20\%$ and $40\%$), two different means ($2$ and $7$) of reliability time for the generating distribution, three different generating distributions (Weibull, gamma, log-normal), and two types of censoring (right and left). In all these cases, we have fixed the variance of the generating distribution to $5$. We have used a non-random censored approach to guarantee the desired proportions of censure, that is, for each scenario, we have fixed a time for which all values that are larger than this time are assumed censored for the right-censored data (series systems), and all values smaller than this fixed time are assumed censored for the left-censored data (parallel systems). In order to improve the analysis, $100$ copies of each scenario were considered.

To summarize and to compare the results of the simulation, we have evaluated the bias and the mean squared error (MSE) of the estimated mean reliability time, for each scenario. The results are presented in Tables \ref{tab_series_weib}-\ref{tab_parallel_lnorm}. One can see from the results that all biases and MSEs are close to zero, indicating that, in all scenarios, the model has estimated well the true mean reliability time. Even in the most challenging scenarios, which are those with small sample size ($n = 30$) and large proportion of censoring ($40\%$), the estimated mean reliability time has had a good performance. When the generating distribution is Weibull, the model should obviously estimate well, yet the results show that the performances were good even for the other two generating (non-Weibull) distributions. 

\begin{table}[H]
\centering
\caption{\label{tab_series_weib} Bias and mean squared error (MSE) of the $\Eset(T)$ estimate for data generated from the Weibull distribution with right-censored data.} 
\begin{tabular}{cccccccccc} 
\hline 
& & \multicolumn{2}{c}{n = 30} & \multicolumn{2}{c}{n = 100} & \multicolumn{2}{c}{n = 1000} \\ 
Cens. & True $\Eset(T)$ &    Bias & MSE     &    Bias &     MSE &    Bias & MSE  \\ 
\hline 
   0\% &          2 & -0.1300 & ~0.2217 & -0.0426 & ~0.0572 & -0.0057 & ~0.0043 \\ 
   0\% &          7 & ~0.0293 & ~0.1523 & ~0.0026 & ~0.0474 & -0.0095 & ~0.0043 \\ 
  20\% &          2 & -0.2813 & ~0.4893 & -0.1242 & ~0.1275 & -0.0200 & ~0.0068 \\ 
  20\% &          7 & -0.0522 & ~0.2307 & -0.0233 & ~0.0565 & -0.0148 & ~0.0047 \\ 
  40\% &          2 & -0.4293 & ~0.7364 & -0.2394 & ~0.2925 & -0.0239 & ~0.0108 \\ 
  40\% &          7 & -0.1773 & ~0.3590 & -0.0661 & ~0.0874 & -0.0162 & ~0.0055 \\ 
\hline 
\end{tabular} 
\end{table} 

\begin{table}[H]
\centering
\caption{\label{tab_series_gamma} Bias and mean squared error (MSE) of the $\Eset(T)$ estimate for data generated from the gamma distribution with right-censored data.} 
\begin{tabular}{cccccccccc} 
\hline 
& & \multicolumn{2}{c}{n = 30} & \multicolumn{2}{c}{n = 100} & \multicolumn{2}{c}{n = 1000} \\ 
Cens. & True $\Eset(T)$ &    Bias & MSE     &    Bias &     MSE &    Bias & MSE  \\ 
\hline 
   0\% &          2 & -0.1407 & ~0.2286 & -0.0191 & ~0.0593 & -0.0003 & ~0.0055 \\ 
   0\% &          7 & ~0.0006 & ~0.1845 & ~0.0081 & ~0.0623 & ~0.0098 & ~0.0048 \\ 
  20\% &          2 & -0.3829 & ~0.6346 & -0.1132 & ~0.1234 & -0.0560 & ~0.0111 \\ 
  20\% &          7 & ~0.0569 & ~0.1721 & ~0.1121 & ~0.0669 & ~0.1049 & ~0.0153 \\ 
  40\% &          2 & -0.5068 & ~0.9424 & -0.3084 & ~0.3807 & -0.1174 & ~0.0299 \\ 
  40\% &          7 & ~0.1389 & ~0.2505 & ~0.2372 & ~0.1107 & ~0.2336 & ~0.0582 \\ 
\hline 
\end{tabular} 
\end{table} 

\begin{table}[H]
\centering
\caption{\label{tab_series_lnorm} Bias and mean squared error (MSE) of the $\Eset(T)$ estimate for data generated from the log-normal distribution with right-censored data.} 
\begin{tabular}{cccccccccc} 
\hline 
& & \multicolumn{2}{c}{n = 30} & \multicolumn{2}{c}{n = 100} & \multicolumn{2}{c}{n = 1000} \\ 
Cens. & True $\Eset(T)$ &    Bias & MSE     &    Bias &     MSE &    Bias & MSE  \\ 
\hline 
   0\% &          2 & -0.1160 & ~0.1757 & -0.0024 & ~0.0396 & -0.0089 & ~0.0050 \\ 
   0\% &          7 & -0.0348 & ~0.1551 & ~0.0462 & ~0.0429 & ~0.0220 & ~0.0052 \\ 
  20\% &          2 & ~0.0720 & ~0.1410 & ~0.2475 & ~0.0779 & ~0.2523 & ~0.0655 \\ 
  20\% &          7 & ~0.0786 & ~0.1492 & ~0.1848 & ~0.0637 & ~0.1620 & ~0.0296 \\ 
  40\% &          2 & ~0.2373 & ~0.1795 & ~0.4205 & ~0.1981 & ~0.4371 & ~0.1928 \\ 
  40\% &          7 & ~0.2190 & ~0.2094 & ~0.3438 & ~0.1532 & ~0.3337 & ~0.1147 \\ 
\hline 
\end{tabular} 
\end{table} 

\begin{table}[H]
\centering
\caption{\label{tab_parallel_weib} Bias and mean squared error (MSE) of the $\Eset(T)$ estimate for data generated from the Weibull distribution with left-censored data.} 
\begin{tabular}{cccccccccc} 
\hline 
& & \multicolumn{2}{c}{n = 30} & \multicolumn{2}{c}{n = 100} & \multicolumn{2}{c}{n = 1000} \\ 
Cens. & True $\Eset(T)$ &    Bias & MSE     &    Bias &     MSE &    Bias & MSE  \\ 
\hline 
   0\% &          2 & -0.1300 & ~0.2217 & -0.0426 & ~0.0572 & -0.0057 & ~0.0043 \\ 
   0\% &          7 & ~0.0293 & ~0.1523 & ~0.0026 & ~0.0474 & -0.0095 & ~0.0043 \\ 
  20\% &          2 & ~0.1030 & ~0.1178 & ~0.0034 & ~0.0475 & -0.0060 & ~0.0043 \\ 
  20\% &          7 & ~0.0387 & ~0.1562 & -0.0001 & ~0.0490 & -0.0094 & ~0.0043 \\ 
  40\% &          2 & -0.1813 & ~0.2660 & -0.0471 & ~0.0579 & -0.0065 & ~0.0043 \\ 
  40\% &          7 & ~0.0301 & ~0.1813 & ~0.0040 & ~0.0597 & -0.0099 & ~0.0048 \\ 
\hline 
\end{tabular} 
\end{table} 

\begin{table}[H]
\centering
\caption{\label{tab_parallel_gamma} Bias and mean squared error (MSE) of the $\Eset(T)$ estimate for data generated from the gamma distribution with left-censored data.} 
\begin{tabular}{cccccccccc} 
\hline 
& & \multicolumn{2}{c}{n = 30} & \multicolumn{2}{c}{n = 100} & \multicolumn{2}{c}{n = 1000} \\ 
Cens. & True $\Eset(T)$ &    Bias & MSE     &    Bias &     MSE &    Bias & MSE  \\ 
\hline 
   0\% &          2 & -0.1407 & ~0.2286 & -0.0192 & ~0.0594 & -0.0003 & ~0.0055 \\ 
   0\% &          7 & ~0.0006 & ~0.1845 & ~0.0081 & ~0.0623 & ~0.0098 & ~0.0048 \\ 
  20\% &          2 & ~0.1029 & ~0.1038 & ~0.0448 & ~0.0490 & ~0.0008 & ~0.0054 \\ 
  20\% &          7 & ~0.1108 & ~0.2143 & ~0.0950 & ~0.0739 & ~0.0932 & ~0.0139 \\ 
  40\% &          2 & -0.1934 & ~0.2463 & -0.0245 & ~0.0602 & -0.0029 & ~0.0055 \\ 
  40\% &          7 & ~0.2185 & ~0.2842 & ~0.2105 & ~0.1242 & ~0.1850 & ~0.0402 \\ 
\hline 
\end{tabular} 
\end{table} 

\begin{table}[H]
\centering
\caption{\label{tab_parallel_lnorm} Bias and mean squared error (MSE) of the $\Eset(T)$ estimate for data generated from the log-normal distribution with left-censored data.} 
\begin{tabular}{cccccccccc} 
\hline 
& & \multicolumn{2}{c}{n = 30} & \multicolumn{2}{c}{n = 100} & \multicolumn{2}{c}{n = 1000} \\ 
Cens. & True $\Eset(T)$ &    Bias & MSE     &    Bias &     MSE &    Bias & MSE  \\ 
\hline 
   0\% &          2 & -0.1160 & ~0.1757 & -0.0024 & ~0.0396 & -0.0089 & ~0.0050 \\ 
   0\% &          7 & -0.0348 & ~0.1551 & ~0.0462 & ~0.0429 & ~0.0220 & ~0.0052 \\ 
  20\% &          2 & -0.0092 & ~0.1253 & ~0.0254 & ~0.0405 & ~0.0211 & ~0.0052 \\ 
  20\% &          7 & ~0.1003 & ~0.1701 & ~0.1757 & ~0.0784 & ~0.1375 & ~0.0240 \\ 
  40\% &          2 & -0.1165 & ~0.1937 & ~0.0575 & ~0.0441 & ~0.0624 & ~0.0086 \\ 
  40\% &          7 & ~0.2359 & ~0.2727 & ~0.3138 & ~0.1548 & ~0.2763 & ~0.0825 \\ 
\hline 
\end{tabular} 
\end{table} 

\end{example}

\section{Final Remarks}
\label{sec:final}

We have introduced a Bayesian reliability statistical analysis using hierarchical models for the problem of estimating the reliability functions and credible bounds of series and parallel systems. The MCMC has shown good performance in terms of convergence, making the inference process simple and efficient. It shall be noted that this performance is not dependent on our choice of a ``non-informative'' scheme to define the prior hyper-parameters. This is important because other researchers may want to fairly compare our method with other frequentist estimators. However, informative priors may very well produce additional improvements in the estimates. The Example \ref{ex_1} has shown good robustness in the sense that the model has performed well for all components in both series and parallel systems. Another important aspect is that we can obtain credible bounds for the reliability function, task that is usually hard if one uses a plug-in estimator for the reliability function. The Example \ref{ex_2} provides an extensive simulation study with more than one hundred different scenarios. Overall, the model has performed very well for estimating the mean reliability time. Some open questions that should be addressed in future works are  the development of hypothesis tests for the components, for instance, one can have interest in testing the hypothesis of equal means of all components (or a subset of components), and the extension of these ideas to more general coherent systems.

\bibliographystyle{plainnat}

\begin{thebibliography}{9}
\providecommand{\natexlab}[1]{#1}
\providecommand{\url}[1]{\texttt{#1}}
\expandafter\ifx\csname urlstyle\endcsname\relax
  \providecommand{\doi}[1]{doi: #1}\else
  \providecommand{\doi}{doi: \begingroup \urlstyle{rm}\Url}\fi

\bibitem[Bhattacharya and Samaniego(2010)]{Bhattacharya2010}
D.~Bhattacharya and F.~J. Samaniego.
\newblock Estimating component characteristics from system failure-time data.
\newblock \emph{Naval Research Logistics (NRL)}, 57\penalty0 (4):\penalty0
  380--389, 2010.
\newblock \doi{10.1002/nav.20407}.

\bibitem[Bhering et~al.(2012)Bhering, Polpo, and Pereira]{Bhering2012a}
F.~Bhering, A.~Polpo, and C.~A.~B. Pereira.
\newblock A hierarchical {W}eibull {B}ayesian model for series and parallel
  systems.
\newblock \emph{AIP Conference Proceedings}, 1490\penalty0 (1):\penalty0
  59--66, 2012.
\newblock \doi{10.1063/1.4759589}.

\bibitem[Kaplan and Meier(1958)]{KaplanMeier1958}
E.L. Kaplan and P.~Meier.
\newblock Nonparametric estimation from incomplete observations.
\newblock \emph{Journal of the American Statistical Association}, 53:\penalty0
  457--481, 1958.

\bibitem[McLachlan and Krishnan(2008)]{McLachlan2008}
G.~J. McLachlan and T.~Krishnan.
\newblock \emph{The {EM} Algorithm and Extensions}.
\newblock John Wiley \& Sons, 2008.

\bibitem[Polpo and Pereira(2009)]{Polpo2009}
A.~Polpo and C.~A.~B. Pereira.
\newblock Reliability nonparametric {B}ayesian estimation in parallel systems.
\newblock \emph{IEEE Transactions on Reliability}, 58\penalty0 (2):\penalty0
  364--373, June 2009.

\bibitem[Polpo and Sinha(2011)]{Polpo2011}
A.~Polpo and D.~Sinha.
\newblock Correction in {B}ayesian nonparametric estimation in a series system
  or a competing-risks model.
\newblock \emph{Statistics \& Probability Letters}, 81\penalty0 (12):\penalty0
  1756--1759, December 2011.
\newblock \doi{10.1016/j.spl.2011.07.023}.

\bibitem[Polpo et~al.(2009)Polpo, Coque, and Pereira]{Polpo2009b}
A.~Polpo, M.~A. Coque, and C.~A.~B. Pereira.
\newblock Statistical analysis for {W}eibull distributions in presence of right
  and left censoring.
\newblock In \emph{The Proceedings of 8th International Conference on
  Reliability, Maintainability and Safety (ICRMS)}, volume~1, pages 219--223,
  Chengdu, 2009. IEEE.
\newblock \doi{10.1109/ICRMS.2009.5270204}.

\bibitem[Rodrigues et~al.(2012)Rodrigues, Dias, Lauretto, and
  Polpo]{Rodrigues2012}
A.~S. Rodrigues, T.~C.~M. Dias, M.~Lauretto, and A.~Polpo.
\newblock Reliability analysis in series systems: An empirical comparison
  between {B}ayesian and classical estimators.
\newblock In \emph{31st International Workshop on Bayesian Inference and
  Maximum Entropy Methods in Science and Engineering, Waterloo, AIP Conference
  Proceedings}, volume 1443, pages 214--221, Melville, 2012. American Institute
  of Physics.
\newblock \doi{10.1063/1.3703638}.

\bibitem[Salinas-Torres et~al.(2002)Salinas-Torres, Pereira, and
  Tiwari]{Salinas2002}
V.~Salinas-Torres, C.A.B. Pereira, and R.~Tiwari.
\newblock {B}ayesian nonparametric estimation in a series system or a competing
  risks model.
\newblock \emph{Journal of Nonparametric Statistics}, 14:\penalty0 449--458,
  2002.
\end{thebibliography}

\end{document}